%% file: ALPAndPions_arXiv_v2.tex
\documentclass[aps,prl,twocolumn,superscriptaddress,amsfonts,nofootinbib,preprintnumbers]{revtex4-1}
\usepackage{graphicx,epsfig}
\usepackage{ifpdf}
\usepackage{amsmath, amssymb, bm,slashed}
\usepackage{color}
\usepackage{enumerate}

\definecolor{nicered}{rgb}{0.7,0.1,0.1}
\definecolor{nicegreen}{rgb}{0.1,0.5,0.1}

\newcommand{\nn}{\nonumber}

\newcommand{\g}{\gamma}

\makeatletter
\g@addto@macro\bfseries{\boldmath}
\makeatother

\graphicspath{{figures/}}

\usepackage[bookmarks=false]{hyperref}
\hypersetup{colorlinks,citecolor= nicegreen,linkcolor= nicered}

\begin{document}

\def\UCSC{Santa Cruz Institute for Particle Physics, University of California, Santa Cruz, CA 95064, USA}
\def\Berkeley{Ernest Orlando Lawrence Berkeley National Laboratory, University of California, Berkeley, CA 94720, USA}

\title{Constraining axion-like particles from rare pion decays}

\author{Wolfgang Altmannshofer}
\email[Electronic address:]{waltmann@ucsc.edu}
\affiliation{\UCSC}

\author{Stefania  Gori}
\email[Electronic address:]{sgori@ucsc.edu}
\affiliation{\UCSC}

\author{Dean J. Robinson}
\email[Electronic address:]{drobinson@lbl.gov}
\affiliation{\Berkeley}
\affiliation{\UCSC}

\begin{abstract}
Ultraviolet completions for axion-like particles (ALPs) lighter than the neutral pion generically induce ALP-neutral pion mixing, and are therefore sensitive to direct constraints on the mixing angle.
For ALPs below the pion mass, we demonstrate that strong and novel bounds on the ALP-pion mixing angle can be extracted from existing rare pion decay data, measured by the PIENU and PIBETA experiments. 
\end{abstract}

\maketitle

\section{Introduction}
Searches for axion-like particles (ALPs) provide a powerful probe of various extensions of the Standard Model (SM), including models of dark matter, baryogenesis, 
and the strong CP problem (see e.g.~\cite{Jaeckel:2010ni,Jeong:2018jqe,Agrawal:2017ksf,Alves:2017avw}).
In the MeV--GeV mass range, the strongest known constraints arise mainly through their coupling to photons.
At large couplings, bounds from LEP diphoton searches are applicable.
For somewhat smaller couplings, bounds from beam-dump and fixed-target experiments, such as Charm/Nu-Cal, E137, and E141 apply.
Current or future experiments, such as NA62, SeaQuest, and Belle II will further probe this mass--coupling parameter space.

The phenomenology of an ALP, $a$, in this mass range may, however, be partly re-parametrized in terms of its mixing with light unflavored hadrons.
The generic nature of this mixing makes it an attractive phenomenological quantity to explore: UV completions for such ALPs will typically generate ALP-pion mixing, and are therefore sensitive to mixing constraints. 
In this paper, we demonstrate how strong bounds on ALP-pion mixing can be extracted from existing rare pion decay data.

In particular, we leverage the high precision measurements of the chirally-suppressed decay $\pi^+ \to e\nu$ (``$\pi_{e2}$'') 
and of the phase-space suppressed $\pi^+ \to \pi^0 e\nu$ (``$\pi_{\beta}$'') to place strong constraints on $\pi^+ \to a e\nu$ (``$\pi_{a3}$''), as the latter has neither of these suppressions.  
These constraints can in turn be transformed into bounds on ALP-pion mixing
because the $\pi^+ \to a e\nu$ amplitude can be estimated via mixing with the final state $\pi^0$ in $\pi^+ \to \pi^0 e\nu$ (cf. Refs.~\cite{Krauss:1986bq,BARDEEN1987401}).
We derive these bounds for $10\,\text{MeV} \lesssim m_a < m_\pi$, extendible down to the massless limit with improved form factor treatments for the $\pi_{\beta}$ decay.

The PIENU experiment~\cite{PIENU:2011aa,Aguilar-Arevalo:2015cdf} currently provides the highest precision measurement of the $\pi^+ \to e\nu$ branching ratio from decays of stopped charged pions: 
The world average  $\text{Br}[\pi^+ \to e\nu] = (1.230\pm0.004) \times 10^{-4}$~\cite{PhysRevD.98.030001}.
Multicomponent
fits to the measured positron energy spectra have been used in previous studies to 
tightly constrain contributions from heavy sterile neutrino decays, i.e. $\pi^+ \to e N$~\cite{PIENU:2011aa,Aguilar-Arevalo:2017vlf} (see also \cite{Aguilar-Arevalo:2019owf}), 
as well as Majoron-neutrino couplings~\cite{PhysRevD.37.1131}.
In this paper, we derive ALP-pion mixing constraints from PIENU spectra via a similar analysis. 
(The irreducible background from $\text{Br}[\pi^+ \to \pi^0 e\nu]$ is much smaller than the experimental precision.  
This is not the case for e.g. $K \to \pi e \nu$ versus $K \to e\nu$, for which reason we do not study bounds from semileptonic kaon decays.) 

Further, the PIBETA experiment~\cite{Pocanic:2003pf} currently provides the highest precision measurement of the rare $\pi^+ \to \pi^0 e\nu$ decay, $\text{Br}[\pi^+ \to \pi^0 e\nu] = (1.036  \pm 0.006) \times 10^{-8}$, 
including a measurement of the opening angle spectrum of the daughter $\pi^0 \to \gamma\gamma$ process. 
This spectrum has a kinematic edge that is highly sensitive to the $\g\g$ invariant mass. We show that it generates even tighter constraints on $\text{Br}[\pi^+ \to (a \to \gamma\gamma) e\nu]$ for a small $m_a$ range.

Previous analyses have considered bounds on ALP-pion mixing using constraints on $K^+ \to \pi^+ +$\,invisible and estimating the $K^+ \to \pi^+ a$ amplitude from mixing with $K^+ \to \pi^+\pi^0$ (see e.g. Ref.~\cite{Bjorkeroth:2018dzu}).
While powerful, these bounds implicitly require suppression of the ALP-top quark coupling, which can otherwise generate large short-distance $s \to d$ penguin contributions. E.g. in the case of universal ALP-quark coupling,  
the penguins are enhanced by $\frac{m^4_t}{32\pi^4 f_\pi^4} \frac{m_\pi^4}{m_a^4} \frac{|V_{ts}V_{td}|^2}{|V_{us}V_{ud}|^2}\log^2\big[\frac{m_W^2}{m_t^2}\big]$ 
compared to the mixing amplitudes~\cite{Wise:1980ux,Frere:1981cc,Hall:1981bc,BARDEEN1987401}, and naively dominate the $K^+ \to \pi^+ a$ amplitudes.
By contrast, the semileptonic processes we consider arise from tree-level charged-current amplitudes. 
Short-distance contributions are expected to enter only at higher loop and electroweak order, far smaller than the hadron mixing contributions we probe. 
In the context of UV completions, the bounds we derive are therefore independent from kaon bounds.

\section{ALP-Pion mixing}

We consider an ALP, $a$, coupled to SM quarks or gauge bosons, with mass
$m_a < m_\pi$. 
We assume no tree-level ALP-lepton couplings, and consider only the case that the diphoton branching ratio is dominant.

The low energy effective field theory of ALP-SM interactions may be matched onto
the chiral Lagrangian of the light hadrons, such that the ALP-SM interactions involve either mixings with SM hadrons 
or higher-dimension derivatively-coupled interactions to hadrons or gauge bosons (see e.g. Ref.~\cite{Bauer:2017ris}).
In the regime that the ALP-SM effective couplings are perturbative, the physical ALP state
\begin{equation}
	\big| a \big\rangle = (\cos\vartheta + \ldots)\big|a_0\big \rangle + \sin\vartheta\,\big|\pi^0 \big\rangle + \ldots
\end{equation}
in which the angle $\vartheta$ encodes the mixing of the ALP and QCD neutral pion eigenstates, and the ellipsis indicates mixings with other hadrons as allowed by parity and angular momentum conservation.
An amplitude involving a $\pi^0$ generates a contribution to an associated ALP amplitude, via mixing with an off-shell $\pi^0$
\begin{equation}
	\label{eqn:alppiM}
	\langle\ldots |a \ldots\rangle =  \sin\vartheta \langle \ldots | \pi^{*0} \ldots \rangle + \ldots.
\end{equation}
Other contributions may involve mixing with other hadrons, or other UV operators. 

ALP-$\pi^0$ mixing may arise in the chiral Lagrangian at leading order via a mixed kinetic term $\varepsilon \, \partial_\mu a \, \partial^\mu \pi^0$, 
or via a mass mixing term $\mu^2 a \pi^0$.
In UV-complete models, $\varepsilon$ can be generated either through an ALP-gluon coupling or through an ALP coupling to light quarks 
(see e.g. Ref.~\cite{Bauer:2017ris}). Typically $\varepsilon \sim f_\pi/f_a$, where $f_\pi$ ($f_a$) is the pion (ALP) decay constant, 
and in the limit $\varepsilon \ll 1$, $\sin\vartheta \simeq m_a^2 \varepsilon/ (m_\pi^2 - m_a^2)$. 
A large mixing, $\sin\vartheta \lesssim 1$, with $m_a \ll m_{\pi}$ requires $f_a \lesssim f_\pi$, which may be difficult to UV-complete.
By contrast, a Higgs Yukawa-like term $y (a/f_a) \bar{Q}_L H \gamma^5 D_R$ may generate a mass mixing term $\mu^2 \sim y v_{\text{EW}} 4 \pi f_\pi^2/f_a$. 
In the limit $\mu^2 \ll m_\pi^2$, $\sin\vartheta \simeq \mu^2/ (m_\pi^2 - m_a^2)$. 
In this case, even for large mixing $\sin\vartheta \lesssim 1$, one may have $f_a \gg v_{\text{EW}}$,  with $m_a$ remaining arbitrarily small, and corrections to $m_\pi$ negligible.
Thus plausible UV-completions may exist that cover the entire $m_a$--$\sin\vartheta$ plane that we consider in this work.
Other sources of isospin-breaking may generate additional mass mixing terms, that further modify $\sin\vartheta$. 
Hereafter, we shall treat $\sin\vartheta$ as a purely phenomenological mixing parameter -- keeping in mind that it may be re-expressed in terms of UV quantities in a model-dependent way -- and seek to develop direct $\sin\vartheta$ constraints.

\section{ALP lifetime}

The amplitude for the diphoton mode $a \to \g\g$ -- the dominant decay mode for $m_a < m_\pi$ -- presents a simple manifestation of
Eq.~\eqref{eqn:alppiM}: It always receives a contribution from ALP-pion mixing 
$\langle \g\g | a \rangle = \langle \g\g | \pi^{*0} \rangle \langle \pi^{*0} | a \rangle + \ldots=\langle \g\g | \pi^{*0} \rangle \sin\vartheta+ \ldots$, 
with possibly 
additional model-dependent UV contributions from a direct coupling to photons, $g_{a\g}a F_{\mu\nu}  \tilde{F}^{\mu\nu} $ ($\tilde{F}_{\mu\nu} = \varepsilon_{\mu\nu\rho\sigma} F^{\rho\sigma}$).
The diphoton width is then (choosing $f_\pi = 130$\,MeV)
\begin{equation}
	\label{eqn:Gagg}
	\Gamma_{a\g\g} = (g^{\text{eff}}_{a\g})^2 m_a^3/\pi\,,\qquad g^{\text{eff}}_{a\g} = \sin\vartheta \, g_{\pi \g} + g_{a\g}\,,
\end{equation}
with an effective coupling, $g^{\text{eff}}_{a\g}$, and the coupling of the pion to photons is $g_{\pi \g} = \sqrt{2} \alpha/8\pi f_\pi \simeq 3.2 \times 10^{-3}/$\,GeV.

The limits we explore from PIENU and PIBETA data are sensitive to the ALP lifetime, as it determines whether the ALP is prompt or invisible at detector scales.
Since the lifetime~\eqref{eqn:Gagg} is in general independent from $\sin\vartheta$, for the purposes of setting $\sin\vartheta$ limits we shall explore two lifetime regimes:
\begin{enumerate}[i)]
	\item The \emph{prompt} regime, i.e. $g^{\text{eff}}_{a\g}$ is sufficiently large for the ALP to decay within the timing/displacement resolution of the detector, possibly via a large $g_{a\g}$,
	\item The \emph{invisible} regime, i.e. $g^{\text{eff}}_{a\g}$ 
	is sufficiently small for the ALP to be long-lived enough to escape the detector,
	possibly via tuning of $g_{a\g}$ against the mixing contribution.
\end{enumerate}
In this context, we will also consider a pure \emph{mixing scenario}, arising from particularly predictive models that do not contain a UV-contribution to the ALP-photon coupling, such that $g^{\text{eff}}_{a\g} \simeq  \sin\vartheta \, g_{\pi \g}$. 

\section{Pion semileptonic decays to ALPs}

At tree-level, the $\pi^+ \to a e\nu$ parton-level amplitude 
\begin{equation}
{\mathcal{A}[\pi^+ \to a e \nu] \sim}~~~~~~~~~~~~~~~~~~~~~~~~~~~~~~~~~~~~~~~~~~~~~~ \end{equation}
	\vspace{-0.6cm}\begin{figure}[h!]
	\includegraphics[height= 2.5cm]{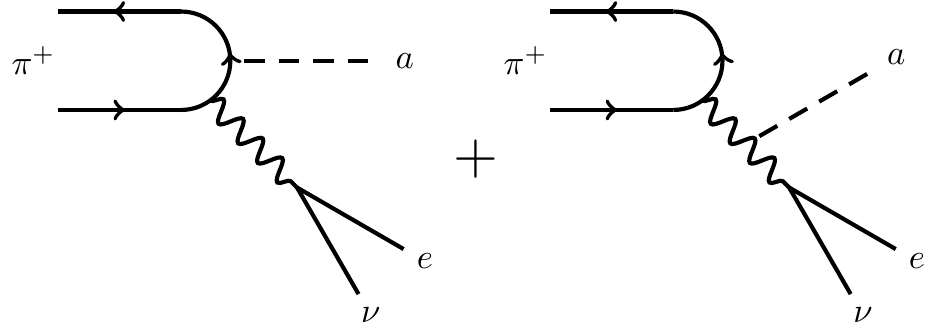}
	\end{figure}

The second term is electroweak suppressed compared to the first, and can be neglected. 
The first term contains the ALP-hadron matrix element of the form
\begin{align}
	\mathcal{A}^\mu \simeq	& \sum_{M^0} \langle a | M^0 \rangle \langle M^0 | \bar{d} \g^\mu u | \pi^+ \rangle \nn \\
		& \quad + \sum_{M^+} \langle 0 | \bar{d} \g^\mu u | M^+ \rangle\langle M^+  a | \bar{q} \slashed{p}_a \g^5 q |\pi^+ \rangle\,, \label{eqn:MEs}
\end{align}
where $M^{0}$ and $M^+$ span complete sets of (multi)hadronic states, with appropriate quantum numbers. 
The axial vector matrix element $\langle M^{0} |\bar d \g^\mu \g^5 u | \pi^+ \rangle$ vanishes by parity and angular momentum conservation.

The leading chirally-unsuppressed contribution to the second term of Eq.~\eqref{eqn:MEs} arises from virtual $\rho^*$ exchange,
and is therefore suppressed by $m_a^2/m_\rho^2$.
The dominant contribution to the matrix element is then generated via off-shell $\pi^{*0}$ `insertion' in the first term, as in Eq.~\eqref{eqn:alppiM}, so that the $\pi^+ \to a$ amplitude
\begin{equation}
	\label{eqn:ampl_approx}
	\mathcal{A}^\mu \simeq \langle a | \pi^{*0} \rangle \langle \pi^{*0} | \bar{d} \g^\mu u | \pi^+ \rangle \equiv \sin\vartheta \langle \pi^{*0} | \bar{d} \g^\mu u | \pi^+ \rangle\,.
\end{equation}

The $\pi^+\to\pi^0 e\nu$ decay is conventionally computed by applying the conserved vector current hypothesis and by mapping to the $\mu \to e \nu\nu$ process, see e.g. Ref.~\cite{RevModPhys.50.573}.
Our estimates for $\pi^+ \to \pi^{*0} e \nu$ will instead be informed by the similar $K^+ \to \pi^0 e \nu$ process, using the language of form factors.
This is a similar, but more general, approach to that of Refs.~\cite{Krauss:1986bq,BARDEEN1987401}, that studied $\pi^+ \to (a \to ee)e\nu$ in the context of the long-defunct $1.8$\,MeV axion anomaly~\cite{PhysRevLett.56.444}.

The hadronic $\pi^+ \to \pi^{*0}$ SM matrix element may be represented by form factors, defined 
via
\begin{align}
	\langle \pi^{*0} |\bar d \g^\mu u | \pi^+ \rangle  & = c_\pi\!\Big[ f_+(p^\mu_+ + p^\mu_0) + (f_0 \!-\!f_+) \frac{m_{+}^2\!-\! m_0^2}{q^2} q^\mu \Big], \nn
\end{align}
in which $q = p_+ - p_0$, the difference of the charged and neutral pseudoscalar momenta, with masses $m_+$ and $m_0$, respectively.
We have defined dimensionless form factors $f_{+,0} = f_{+,0}(q^2)$, such that $f_0$ couples only to the lepton mass. 
Here $c_\pi = 2 \times 1/\sqrt{2}$ is a coupling combinatoric factor multiplied by a Clebsch-Gordan coefficient. (For $K^+ \to \pi^0e\nu$, $c_K = 1/\sqrt{2}$.) 
In the regime $m_+ - m_0 \gg m_e$, the electron mass terms may be neglected, so that the $\pi^+ \to \pi^{*0} e\nu$ rate
\begin{align}
	\frac{d\Gamma}{d q^2}  &= \frac{G_F^2 |V_{ud}|^2 c_\pi^2 m_{+}^3}{24 \pi^3} \bigg(\frac{q^2 - m_e^2}{q^2}\bigg)^2 r \sqrt{w^2 -1} \label{eqn:diffrate}\\
	& \quad  \times\bigg[ f_+^2 r^2 (w^2-1)\bigg(1 + \frac{m_e^2}{2q^2}\bigg) + \frac{3 m_e^2}{8 q^2} f_0^2(1-r^2)^2\bigg]\,, \nn
\end{align}
in which $r = m_0/m_{+}$, the recoil parameter $w = (m_{+}^2 + m_0^2 - q^2)/(2m_{+} m_0)$, 
with range $1 \le w \le  (1 + r^2)/(2r)$, and we neglect small electroweak corrections~\cite{RevModPhys.50.573}. 

Following from the Ademollo-Gatto theorem~\cite{PhysRevLett.4.186,PhysRevLett.13.264, Leutwyler1984}, one expects $f_+(q^2 =0) \simeq 1$ up to corrections that are expected to scale  as $\sim (m^2_{+} - m_0^2)^2/\Lambda_{\text{QCD}}^4$.
The matrix element may be expressed as an analytic function of a conformal expansion parameter $z = (\sqrt{w+1} - \sqrt{2})/(\sqrt{w+1} + \sqrt{2})$~\cite{BOURRELY1981157},
so that provided $|z| \ll 1$, the form factor should be approximately linear in $w$ or $q^2$. 
In the analogous $K^+ \to \pi^0 e \nu$ system, $f_+^K(q^2)$ is well-approximated by a linear function from $f(q^2=0) \simeq 1$ to $f(q^2 = q^2_{\text{max}}) \sim 1.2$, and $|z|_{\text{max}} \simeq 0.098$. 
Thus requiring a sufficiently small $z$, say $|z|_{\text{max}} \lesssim 0.3$ -- equivalent to $r \gtrsim 0.1$ or $m_0\gtrsim 10$ MeV -- and approximating
\begin{equation}
	\label{eqn:AGfp}
	f_+(q^2) \simeq 1\,,
\end{equation}
should provide a lower bound on $f_+$, yielding a conservative estimate for the $\pi^+ \to a e \nu$ rate up to  $\mathcal{O}(10\%)$ uncertainties. 
(In the massless positron limit, applying the approximation~\eqref{eqn:AGfp} to Eq.~\eqref{eqn:diffrate} yields a partial width in agreement with e.g. Eq.~(1) of Ref.~\cite{Pocanic:2003pf} or Eq.~(7.12) of Ref.~\cite{RevModPhys.50.573} to $\mathcal{O}[(1-r)^8]$.) 

\begin{figure}[t!]
	\includegraphics[height= 3.5cm]{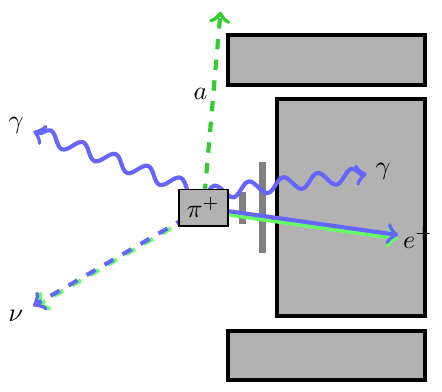}\hfill
	\includegraphics[height= 3.5cm]{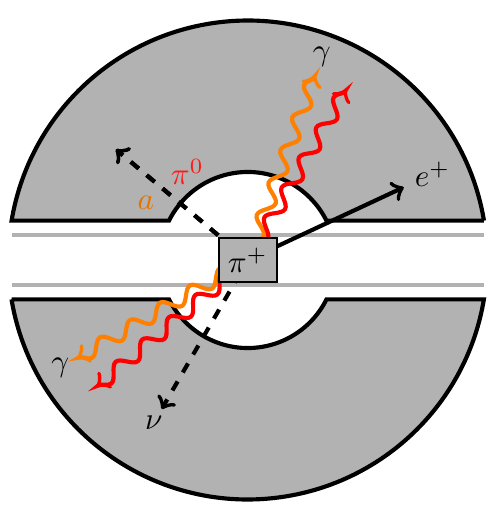}
	\caption{\emph{Left:} Schematic PIENU detector configuration of the target, tracking and calorimeter elements (gray). Overlaid are typical event topologies for the prompt (blue) and invisible (green) ALP scenarios.
	\emph{Right:} Schematic cross-section of the PIBETA detector configuration including the target, tracking and calorimeter elements (gray). Overlaid are typical event topologies at the minimum truth-level opening angle configuration of a $\pi^0$ (red) and lighter prompt ALP (orange) diphoton decay.} 
	\label{fig:PIENUPIBETAdet}
\end{figure}

Combining Eqs.~\eqref{eqn:ampl_approx}--\eqref{eqn:AGfp} with the $\pi^+ \to \ell \nu$ partial width, one obtains the ratio of branching ratios

\begin{equation}
	\frac{\text{Br}[\pi_{a3}]}{\text{Br}[\pi_{\ell2}]} \simeq  \frac{2}{3\pi^2}\frac{c_\pi^2 m_\pi^4 \sin^2\vartheta}{f_\pi^2 m_\ell^2 (1 - m_\ell^2/m_{+}^2)^2} \int_1^{(1+r^2)/2r} \hspace{-1.2cm}r^4(w^2-1)^{3/2} dw\,.\label{eqn:BRrs}
\end{equation}
Using Eq.~\eqref{eqn:BRrs}, we proceed to set bounds on $\sin\vartheta$ from rare pion decay data. 
These bounds rely, in part, on fits to the positron energy spectrum in the parent rest frame. 
At truth level, the positron energy is bounded by $0 \le E_e \le m_+(1-r^2)/2$, and
\begin{equation} 
	\frac{d\Gamma[\pi_{a3}]}{dE_e} = \frac{c_\pi^2 G_F^2  m_+ E_e^2}{8\pi^3} \frac{\big((1 -r^2)m_+ - 2E_e\big)^2}{m_+ - 2E_e}\,.
\end{equation}

\section{PIENU Residuals Bound}
The PIENU experiment~\cite{Aguilar-Arevalo:2015cdf} measures the $\pi^+ \to e\nu$ branching ratio from a sample of stopped pions, 
by determining the positron yield in the electromagnetic (EM) inclusive decay $\pi^+ \to e\nu(\gamma)$ compared to the cascade $\pi^+ \to (\mu \to e\nu\nu)\nu(\gamma)$.
The main experimental components comprise a target, silicon strips and wire chambers for high precision tracking, a positron calorimeter to reconstruct the positron energy, 
and a semi-hermetic calorimeter array to capture EM showers. The combined calorimeter energy is given by the sum of positron energy and EM showers, $E_{\text{cal}} = E_{e} + E_{\text{EM}}$.
A sketch of the PIENU detector is shown on the left in Fig.~\ref{fig:PIENUPIBETAdet}.

The relevant backgrounds include not only the $\pi^+ \to \mu \to e$ cascade, but also contributions from pion decays-in-flight, stopped muon decays, and radiative $\mu$ decays to energetic photons. 
Their branching ratios overwhelmingly dominate the signal mode.
Timing cuts are used to suppress these large backgrounds compared to the prompt $\pi \to e\nu(\gamma)$ modes. 
A simultaneous fit of the timing distributions for both signal and backgrounds then permits measurement of the ratio, 
$R_{e/\mu} = \Gamma[\pi \to e \nu(\gamma)]/\Gamma[\pi \to \mu \nu(\gamma)]$, at the $10^{-3}$ level, from which the $\pi^+ \to e\nu$ branching ratio is inferred.

\begin{figure}[t!]
	\includegraphics[width = 0.9\linewidth]{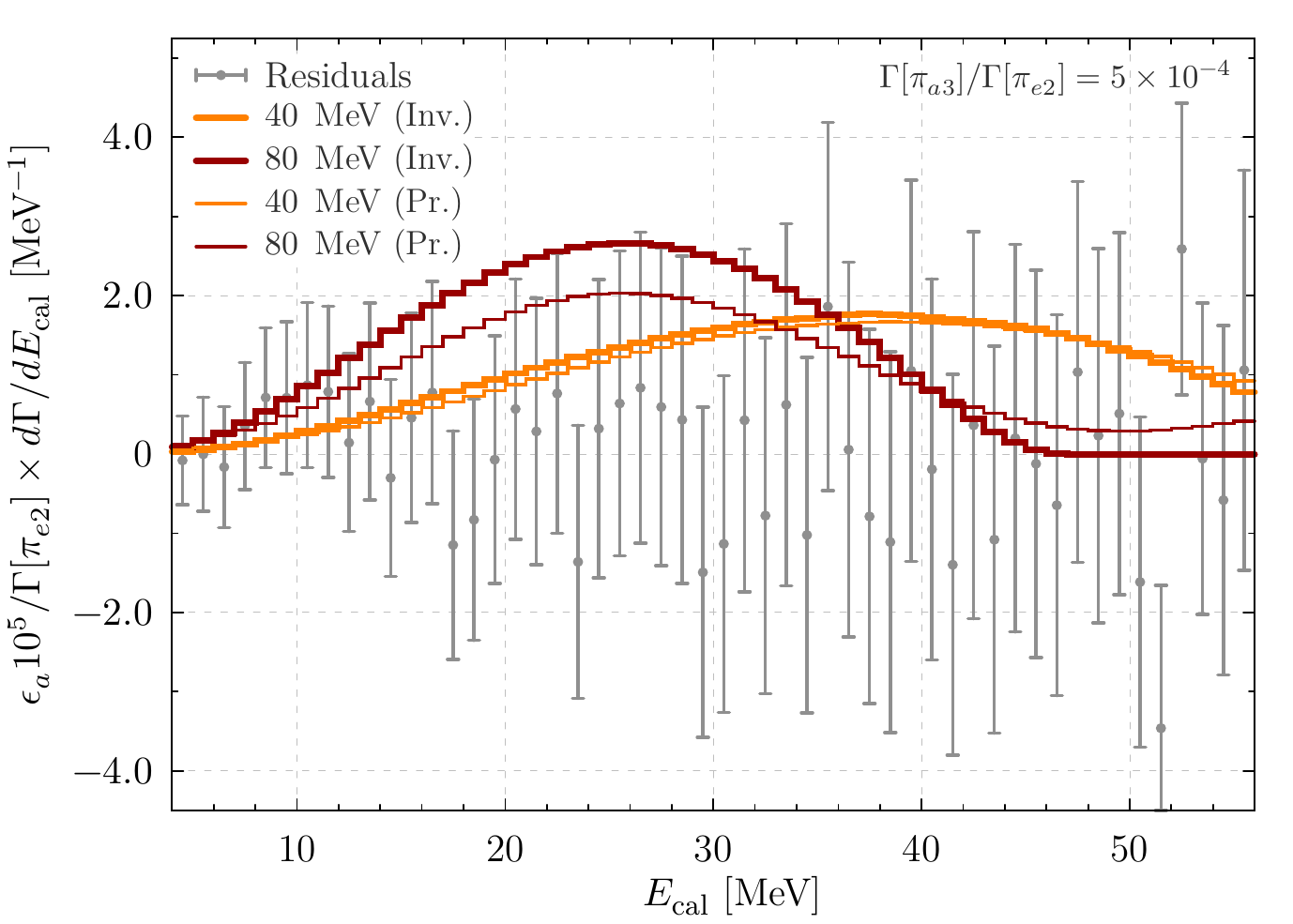}
	\caption{PIENU fit bin residuals (grey) for the $E_{\text{cal}}$ distribution in the low-energy regime, normalized against the $\pi^+ \to e \nu$ (``$\pi_{e2}$'') rate. Overlaid are $\pi^+ \to a e \nu$ (``$\pi_{a3}$'') binned spectra for the prompt (thin solid) and invisible (thick solid) regimes, with $m_a = 40$\,MeV (orange) and $80$\,MeV (red). The spectra include acceptance corrections, with total acceptance $\epsilon_a$, but are normalized such that $\Gamma[\pi_{a3}]/\Gamma[\pi_{e2}] = 5\times10^{-4}$.} 
	\label{fig:PIENUres}
\end{figure}

\begin{figure*}[t!]
	\includegraphics[width = 0.487\linewidth]{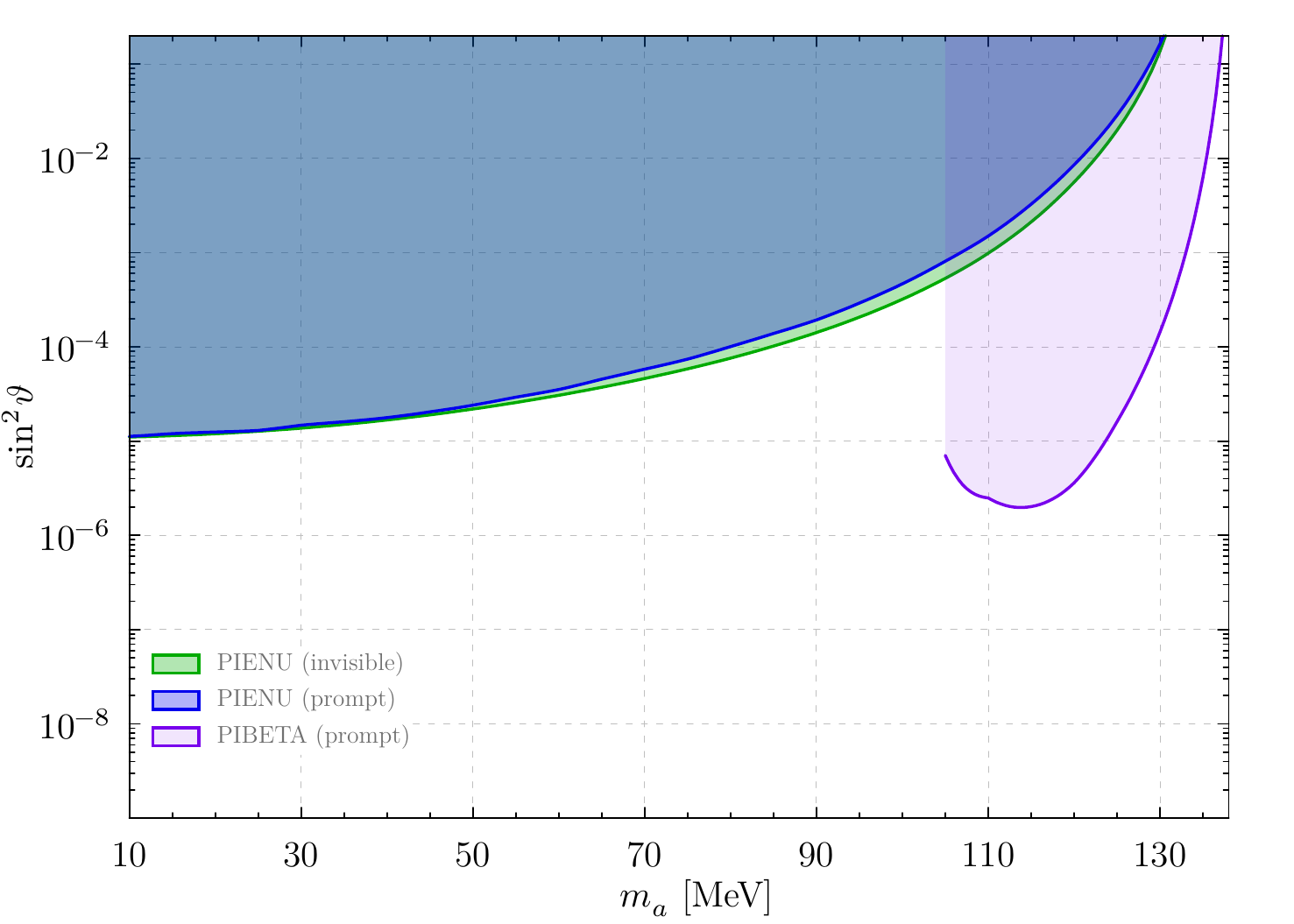}\hfill
	\includegraphics[width = 0.511\linewidth]{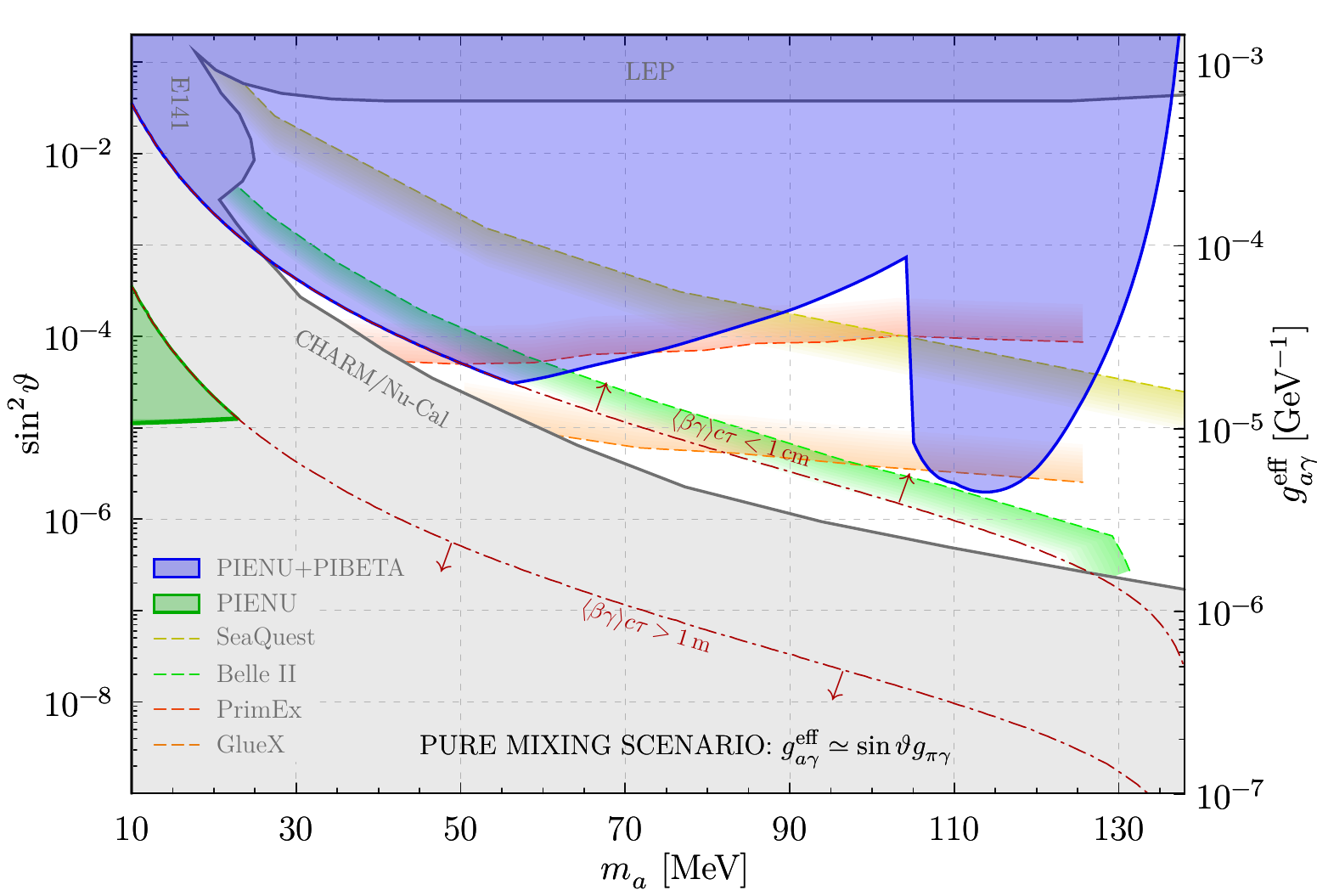}
	\caption{\emph{Left:} 95\% CL exclusion regions from the PIENU fit residuals for the prompt (blue) and invisible (green) ALP regimes.
	Also shown is the PIBETA exclusion from the $\theta_{\g\g}$ spectrum for prompt ALPs (purple). 
	\emph{Right:} Combined  PIENU and PIBETA exclusion regions for the models with $g^{\text{eff}}_{a\g} \simeq  \sin\vartheta \, g_{\pi \g}$ (``pure mixing scenario''). In this case, the mean characteristic decay length regions
	$\langle\beta\gamma\rangle c\tau < 1$\,cm (dot-dashed red line and above) and $\langle\beta\gamma\rangle c\tau > 1$\,m (dot-dashed red line and below) approximately delineate where the prompt and invisible regime exclusions apply, respectively.
	Also shown are exclusions from CHARM/Nu-Cal, E137, E141 and LEP (grey), and projected reaches for SeaQuest (yellow line and below)~\cite{Berlin:2018pwi}, Belle II (green line and below)~\cite{Dolan:2017osp}, PrimEx (red line and above) and GlueX (orange line and above)~\cite{Aloni:2019ruo}.}
	\label{fig:PIENUbnds}
\end{figure*}

The $E_{\text{cal}}$ distribution for the $\pi^+ \to e\nu$ mode is sharply peaked at $(m_{+}^2 + m_e^2)/2m_{+} \simeq 69.8$\,MeV, 
with a low-energy tail arising from EM shower losses.
The (timing-cut-suppressed) backgrounds from the $\pi^+ \to \mu \to e$ cascade or muon decays-in-flight are, by contrast, smoothly distributed in the low-energy region 
$E_{\text{cal}} < E_0 \simeq 52$\,MeV, the endpoint.

Refs.~\cite{PIENU:2011aa,Aguilar-Arevalo:2017vlf} perform a precision fit of the measured $E_{\text{cal}}$ distribution in the low-energy region
to the combination of the (simulated) $\pi^+ \to e\nu(\gamma)$ low-energy tail and the background distributions.
The bin residuals of this fit can be used to place strong constraints on additional prompt contributions from exotic $\pi^+ \to e X$, where $X$ has sufficient invariant mass to push 
the \emph{signal} $E_{\text{cal}}$ distribution into the low-energy fit region.
Refs.~\cite{PIENU:2011aa,Aguilar-Arevalo:2017vlf} consider the case that $X = N$, a heavy sterile neutrino. 
In Fig.~\ref{fig:PIENUres} we show the residuals of Ref.~\cite{Aguilar-Arevalo:2017vlf} used for such an analysis, normalized against the $\pi^+ \to e \nu$ branching ratio.
In this work we consider $X = a\nu$, making use of Eq.~\eqref{eqn:BRrs} to convert the bound on branching ratios to a bound on $\sin\vartheta$. 
(More precise limits will require a dedicated analysis fitting the $X = a\nu$ signal template simultaneously with the background components.)

We characterize whether the ALP is prompt or invisible by considering whether the mean characteristic ALP displacement 
from decay-in-flight, $\langle \beta\gamma \rangle c\tau$, is inside the target or outside the calorimeter radius, respectively: 
We treat the PIENU target size as $\sim 1$\,cm and the calorimeter size as $\sim 1$\,m.
(A full study of regimes outside the prompt or invisible limits requires simulation of the PIENU response when the EM shower is somewhat spatially 
or time-displaced from the prompt decays, but still within the detector acceptance.)

In the invisible ALP regime 
the $E_{\text{cal}}$ distribution receives no additional contributions from $a \to \g\g$.
In Fig.~\ref{fig:PIENUres} we show the corresponding binned positron energy spectra (thick lines) from $\pi^+ \to a e \nu$ decays for $m_a = 40$ and $80$\,MeV, 
including quoted acceptance corrections~\cite{Aguilar-Arevalo:2017vlf}.

In the prompt ALP regime, however, daughter photons of the ALP may contribute to the measured $E_{\text{cal}}$ in the event. 
For $m_a \sim m_\pi$, the ALP is slow enough that one photon may hit the PIENU positron calorimeter within the $\sim 20\%$ positron acceptance~\cite{Aguilar-Arevalo:2017vlf}, as sketched in Fig.~\ref{fig:PIENUPIBETAdet} 
(deposition in the outer calorimeters is required to be $<2$\,MeV~\cite{Aguilar-Arevalo:2017vlf}, thereby excluding hard photon contributions in those).
But for $m_a \ll m_\pi$, the ALP momentum may back-react against the lepton system, such that the daughter photons, which decay in a narrow cone around the ALP momentum, miss the acceptance. 
In Fig.~\ref{fig:PIENUres} we show the corresponding binned positron spectra (thin lines) for the same two mass benchmarks. 
The heavier $80$\,MeV benchmark is slightly altered by a longer tail.

While Ref.~\cite{Aguilar-Arevalo:2017vlf} does not quote the bin residual correlations, one may reproduce quoted $\pi^+ \to e N$ bounds assuming nearby bins are uncorrelated. 
We therefore extend this assumption to treat all bins as uncorrelated over the measured energy range. 
Under this assumption, in the left of Fig.~\ref{fig:PIENUbnds} we show the corresponding 95\% CL exclusion regions in the $\sin^2\vartheta$--$m_a$ parameter space, for both the invisible (green) and prompt (blue) regimes.
The excluded regions in $\sin^2\vartheta$ for the prompt and invisible cases differ at most by $\mathcal{O}(1)$ and extend down to $\sin^2\vartheta\gtrsim 10^{-5}$. 
This corresponds to branching ratios as small as $\mathcal O(10^{-8})$.


Independent of the relationship between $\sin\vartheta$ and the lifetime (cf. Eq.~\eqref{eqn:Gagg}), 
requiring a prompt ALP -- $\langle \beta\gamma \rangle c\tau < 1$\,cm -- directly implies a lower bound on $g^{\text{eff}}_{a\g}$. 
Over the ALP mass ranges considered in this paper, we have checked that this bound is far smaller than the direct $g^{\text{eff}}_{a\g}$ bounds from LEP tri-photon searches \cite{Mimasu:2014nea,Jaeckel:2015jla}. 
Electron fixed-target experiments such as NA64 \cite{Banerjee:2017hhz} and LDMX \cite{Akesson:2018vlm}, as well as Belle(II)~\cite{Dolan:2017osp} and BaBar \cite{delAmoSanchez:2010ac}, also have invisible ALP searches. 
However, for these experiments, the ALP production and lifetime is controlled by $g^{\rm{eff}}_{a\gamma}$, independent of $\sin\vartheta$. 
Hence these constraints do not appear in the left panel of Fig.~\ref{fig:PIENUbnds}. 

The pure mixing scenario ($g_{a\gamma}=0$ in Eq.~\eqref{eqn:Gagg}) fixes the relationship between the $\pi^+ \to a e\nu$ branching ratio and the ALP lifetime, 
and may therefore interpolate between the prompt and invisible regimes in different parts of the $\sin^2\vartheta$--$m_a$ space.
For this scenario, in the right side of Fig.~\ref{fig:PIENUbnds} the region $\langle \beta\gamma \rangle c\tau < 1$\,cm ($>1$\,m) is above (below) the red dot-dashed contours.
Above (below) the $1$\,cm ($1$\,m) contour, the prompt (invisible) exclusion should be a good proxy for the pure mixing scenario.

Further, in the pure mixing scenario, the relation $g_{a\g}^{\text{eff}} \simeq \sin\vartheta \, g_{\pi\g}$ enables recasting of beam-dump, collider, 
and fixed-target experiment bounds on $g^{\text{eff}}_{a\g}$ onto the $\sin^2\vartheta$--$m_a$ space.
For $m_a < m_\pi$, the relevant bounds are set by the CHARM/Nu-Cal \cite{Bergsma:1985qz,Blumlein:1990ay,Blumlein:1991xh,Dobrich:2015jyk}, E137 \cite{Bjorken:1988as}, 
E141 \cite{Dobrich:2017gcm} and LEP \cite{Mimasu:2014nea,Jaeckel:2015jla} experiments, corresponding in the right panel of Fig.~\ref{fig:PIENUbnds} to the gray regions. 
(Roughly scaling the pion interaction length with $\sin^2\vartheta \lesssim 10^{-3}$, the ALP interaction length in matter is naively $\gtrsim 10^3$\,m, far larger than the typical path length in beam dump experiments, so that their constraints continue to apply in the pure mixing scenario.)
We see in Fig.~\ref{fig:PIENUbnds} that the PIENU data places powerful new constraints on ALPs \emph{in the pure mixing scenario} for $m_a \gtrsim 25$\,MeV. 

These  constraints will be complemented in the future by proton fixed-target beam-dump experiments, such as SeaQuest \cite{Berlin:2018pwi} searching for $3\gamma$ signatures, or Belle II monophoton searches~\cite{Dolan:2017osp}.
In the right panel of Fig.~\ref{fig:PIENUbnds}, we show the SeaQuest (yellow, $10^{20}$ protons on target) and Belle II (green) reaches as representatives of experiments 
capable of setting limits in the $\sin^2\vartheta$--$m_a$ space in the pure mixing scenario.
Part of the $\sin^2\vartheta$--$m_a$ space may also be tested by NA62 running in beam-dump mode~\cite{Beacham:2019nyx}, and FASER~\cite{Feng:2018noy}. 
A slightly larger region of parameter space could be probed by SHiP \cite{Alekhin:2015byh},
as well as by PrimEx and GlueX (region above the red and orange lines, respectively)~\cite{Aloni:2019ruo}.

\section{PIBETA Diphoton Bound}
The PIBETA experiment~\cite{Pocanic:2003pf, Frlez:2003vg} measures the rare $\pi^+ \to (\pi^0 \to \g\g)e\nu$ branching ratio from a sample of stopped pions, 
by triggering on the prompt $\pi^0 \to \g\g$ decay in coincidence with a positron track.
The main detector elements relevant here are a near-spherical electromagnetic calorimeter, and cylindrical multi-wire proportional tracking chambers surrounded by plastic scintillator.
A schematic of the experiment is shown on the right in Fig.~\ref{fig:PIENUPIBETAdet}.

The photon showers are required to each have energy $E_\g > m_\mu/2$, beyond the kinematic endpoint of stopped $\mu \to e\nu\nu$ background decays. 
The normalization of the $\pi^+ \to \pi^0 e\nu$ rate is obtained via comparison with a large prescaled sample of non-prompt single positron track events, including both $\pi^+ \to e\nu$ and in-flight $\mu \to e\nu\nu$ backgrounds.
This entails a simultaneous fit of signal and background kinematic and timing distributions.

Reconstruction of the diphoton pair includes measurement of the diphoton opening angle in the lab frame. At truth level, this angle is bounded via
\begin{equation}
	-1 \le \cos\theta_{\g\g} \le -1 + 2\big[(1-r^2)/(1+ r^2)\big]^2\,.
\end{equation}
The maximum (minimum) cosine corresponds to diphoton emission perpendicular (parallel) to the $\pi^+$ direction of flight in the $\pi^0$ rest frame, 
generating a sharp kinematic edge (smooth kinematic endpoint) in the $\theta_{\g\g}$ spectrum.
Because the upper bound increases as $m_a$ decreases, the \emph{prompt} diphoton decay of an ALP in $\pi^+ \to (a\to\g\g)e\nu$ with $m_a < m_\pi$ 
may produce diphoton showers with truth-level opening angles beyond the $\pi^0$ edge at $\sim 176^\circ$.
In Fig.~\ref{fig:PIENUPIBETAdet} we show schematically the maximum truth-level $\cos\theta_{\g\g}$ configuration for a $\pi^0$ compared to a lighter ALP.

\begin{figure}[t!]
	\includegraphics[width = 0.9\linewidth]{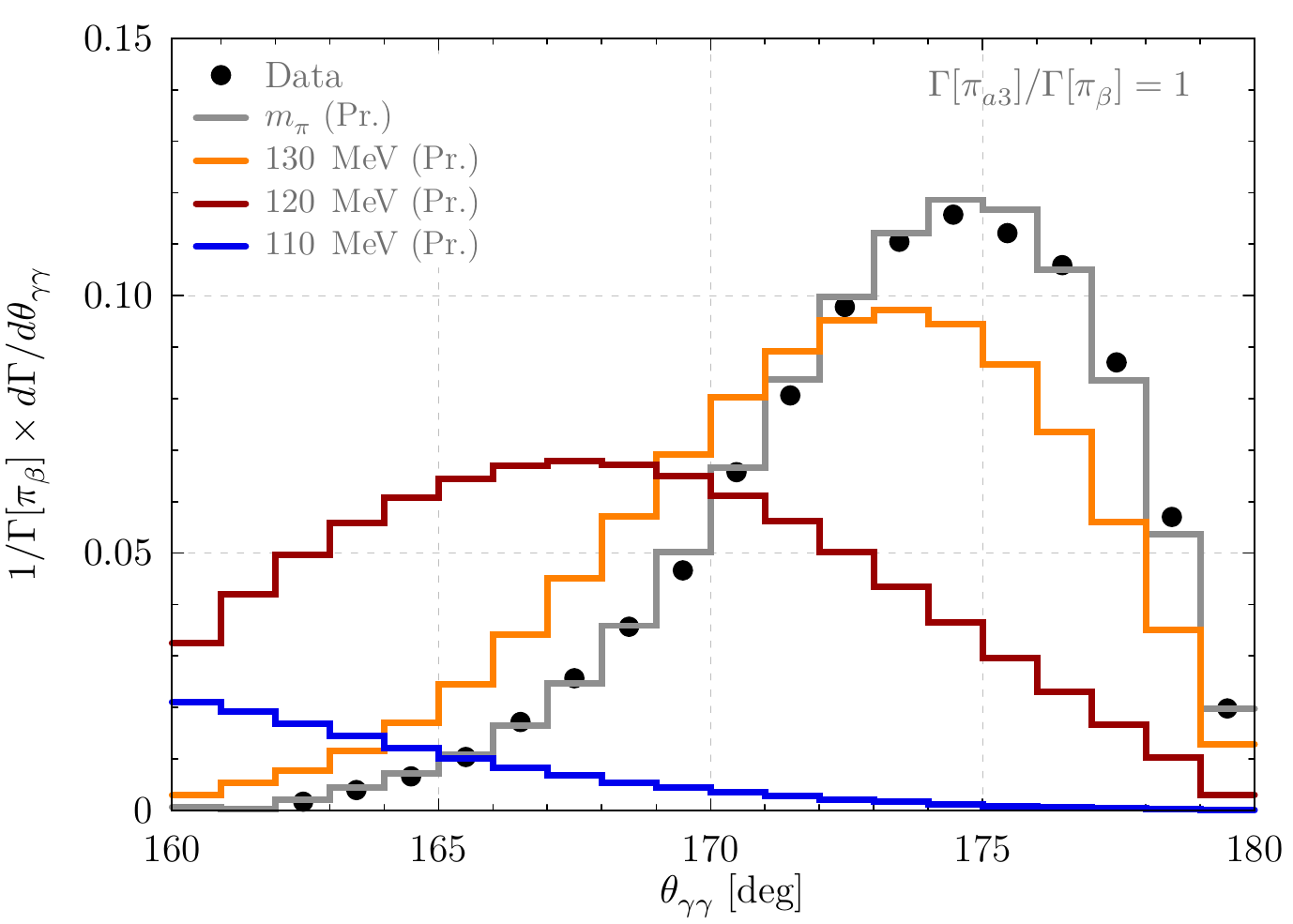}
	\caption{PIBETA reconstructed diphoton opening angle distribution (black) for $\pi^+ \to (\pi^0 \to \g\g)e \nu$, normalized to unity. Also shown are $\pi^+ \to a e\nu$ binned spectra for the prompt regime, with $m_a = 110$, $120$, $130$\,MeV and $m_\pi$. The spectra are normalized such that $\Gamma[\pi^+ \to a e\nu]/\Gamma[\pi^+ \to \pi^0 e \nu] = 1$.}
	\label{fig:PIBETAth}
\end{figure}

In practice, the finite detector-level angular resolution smears out the reconstructed $\theta_{\g\g}$ distribution and thus the $\theta_{\g\g}$ edge. 
For an angular smearing $\sigma_{\theta_{\g\g}} \simeq 2.25^\circ$ and requiring both photons' energy $E_\g > 53$\,MeV~\cite{Frlez:2003vg}, 
we show in Fig.~\ref{fig:PIBETAth} the expected $\theta_{\g\g}$ distributions for several $m_a$ benchmarks, as well as for $\pi^0$,
compared to the measured $\theta_{\g\g}$ spectrum for $160^\circ \le \theta_{\g\g} \le 180^\circ$~\cite{Pocanic:2003pf}. 
The $\pi^0$ spectrum (grey) agrees well with the data. 
For $m_a \lesssim 110$\,MeV, the photon energy cut significantly suppresses the $\theta_{\g\g}$ spectrum in the $160$--$180^\circ$ range.

The PIBETA experiment does not provide residuals for the fit of the simulated $\pi^+\to\pi^0 e\nu$ opening angle spectrum to the data. 
We extract an approximate, estimated bound on $\sin^2\vartheta$, by conservatively requiring that the integrated contribution to the $\theta_{\g\g}$ spectrum in the $160$--$180^\circ$ range from $\pi^+ \to a e\nu$ 
does not exceed the quoted $0.6\%$ uncertainty for the $\pi^+\to\pi^0 e\nu$ branching ratio. 
In Fig.~\ref{fig:PIENUbnds} we show the corresponding exclusion (purple region in the left panel). 
This exclusion will likely be much stronger if the full differential information shown in Fig.~\ref{fig:PIBETAth} can be incorporated. 
This approximate bound from PIBETA data sets the most stringent bound on the mixing angle $\sin\vartheta$ for prompt regime ALPs with masses above $\sim 100$ MeV.
A future data analysis for $\theta_{\gamma\gamma}<160^\circ$ could lead to stringent constraints also for $m_a < 100$\,MeV.

\section{Conclusions and Outlook}
Models for axion-like-particles (ALPs) generically predict mixing between the ALP and the SM neutral pion. We have derived strong new constraints on ALP-pion mixing, by extracting constraints on the $\pi^+ \to a e \nu$ branching ratio from the rare pion decay data measured by the PIENU and PIBETA experiments. 

In the pure mixing scenario, these constraints complement existing exclusions as well as the reaches of planned experiments, 
leading to near complete coverage of the $\sin^2\vartheta$--$m_a$ space over many decades of the mixing angle for $10{\rm{MeV}}\lesssim m_a\lesssim m_\pi$. 
Beyond the pure mixing scenario, the constraints provide exclusions for a wide range of UV ALP models that generate ALP-pion mixing.
Because they arise from charged current tree-level processes, these exclusions can probe UV models that are characteristically different from those probed by similar bounds extracted from $K^+ \to \pi^+ +$\,invisible decays.

Our approximate treatments of the detector responses can be improved by dedicated ALP analyses in future $\pi^+ \to e\nu$ or $\pi^+ \to \pi^0 e \nu$ measurements, 
that account for e.g. bin correlations, effects of displaced ALP decays, and/or make use of other differential information.
Our results rely on theoretical approximations, expected to introduce no more than $\mathcal{O}(10\%)$ uncertainties, that may be improved with more detailed treatments of the $\pi^+ \to \pi^{*0}$ form factors. 
This in turn would permit extension of these bounds to lower ALP masses, below $\sim10$\,MeV.

\section*{Acknowledgements}
We thank Ketevi Assamagan and Emil Frlez for consultations about the PIBETA experiment and Doug Bryman for consultations about the PIENU experiment. 
We thank Bob Cahn, Zoltan Ligeti, Michele Papucci and Mike Williams for helpful discussions.
The research of WA is supported by the National Science Foundation under Grant No. NSF 1912719.
The research of SG is supported by the National Science Foundation under the CAREER grant PHY-1915852.
The work of DR is supported by the U.S.\ Department of Energy under contract DE-AC02-05CH11231,
and was supported in part by the National Science Foundation under Grant No. NSF 1912719.
WA and SG would like to thank KITP for hospitality and acknowledge partial support by the National Science Foundation under Grant No. NSF PHY-1748958. 
SG would like to thank the Aspen Center for Physics, which is supported by National Science Foundation grant PHY-1607611, where this work was completed.

\input{ALPAndPions_arXiv_v2.bbl}

\end{document}

%% file: ALPAndPions_arXiv_v2.bbl
%

%% file: ALPAndPions_arXiv_v2.bbl
\begin{thebibliography}{42}%
\makeatletter
\providecommand \@ifxundefined [1]{%
 \@ifx{#1\undefined}
}%
\providecommand \@ifnum [1]{%
 \ifnum #1\expandafter \@firstoftwo
 \else \expandafter \@secondoftwo
 \fi
}%
\providecommand \@ifx [1]{%
 \ifx #1\expandafter \@firstoftwo
 \else \expandafter \@secondoftwo
 \fi
}%
\providecommand \natexlab [1]{#1}%
\providecommand \enquote  [1]{``#1''}%
\providecommand \bibnamefont  [1]{#1}%
\providecommand \bibfnamefont [1]{#1}%
\providecommand \citenamefont [1]{#1}%
\providecommand \href@noop [0]{\@secondoftwo}%
\providecommand \href [0]{\begingroup \@sanitize@url \@href}%
\providecommand \@href[1]{\@@startlink{#1}\@@href}%
\providecommand \@@href[1]{\endgroup#1\@@endlink}%
\providecommand \@sanitize@url [0]{\catcode `\\12\catcode `\$12\catcode
  `\&12\catcode `\#12\catcode `\^12\catcode `\_12\catcode `\%12\relax}%
\providecommand \@@startlink[1]{}%
\providecommand \@@endlink[0]{}%
\providecommand \url  [0]{\begingroup\@sanitize@url \@url }%
\providecommand \@url [1]{\endgroup\@href {#1}{\urlprefix }}%
\providecommand \urlprefix  [0]{URL }%
\providecommand \Eprint [0]{\href }%
\providecommand \doibase [0]{http://dx.doi.org/}%
\providecommand \selectlanguage [0]{\@gobble}%
\providecommand \bibinfo  [0]{\@secondoftwo}%
\providecommand \bibfield  [0]{\@secondoftwo}%
\providecommand \translation [1]{[#1]}%
\providecommand \BibitemOpen [0]{}%
\providecommand \bibitemStop [0]{}%
\providecommand \bibitemNoStop [0]{.\EOS\space}%
\providecommand \EOS [0]{\spacefactor3000\relax}%
\providecommand \BibitemShut  [1]{\csname bibitem#1\endcsname}%
\let\auto@bib@innerbib\@empty
\bibitem [{\citenamefont {Jaeckel}\ and\ \citenamefont
  {Ringwald}(2010)}]{Jaeckel:2010ni}%
  \BibitemOpen
  \bibfield  {author} {\bibinfo {author} {\bibfnamefont {J.}~\bibnamefont
  {Jaeckel}}\ and\ \bibinfo {author} {\bibfnamefont {A.}~\bibnamefont
  {Ringwald}},\ }\href {\doibase 10.1146/annurev.nucl.012809.104433} {\bibfield
   {journal} {\bibinfo  {journal} {Ann. Rev. Nucl. Part. Sci.}\ }\textbf
  {\bibinfo {volume} {60}},\ \bibinfo {pages} {405} (\bibinfo {year} {2010})},\
  \Eprint {http://arxiv.org/abs/1002.0329} {arXiv:1002.0329 [hep-ph]}
  \BibitemShut {NoStop}%
\bibitem [{\citenamefont {Jeong}\ \emph {et~al.}(2018)\citenamefont {Jeong},
  \citenamefont {Jung},\ and\ \citenamefont {Shin}}]{Jeong:2018jqe}%
  \BibitemOpen
  \bibfield  {author} {\bibinfo {author} {\bibfnamefont {K.~S.}\ \bibnamefont
  {Jeong}}, \bibinfo {author} {\bibfnamefont {T.~H.}\ \bibnamefont {Jung}}, \
  and\ \bibinfo {author} {\bibfnamefont {C.~S.}\ \bibnamefont {Shin}},\
  }\href@noop {} {\  (\bibinfo {year} {2018})},\ \Eprint
  {http://arxiv.org/abs/1811.03294} {arXiv:1811.03294 [hep-ph]} \BibitemShut
  {NoStop}%
\bibitem [{\citenamefont {Agrawal}\ and\ \citenamefont
  {Howe}(2018)}]{Agrawal:2017ksf}%
  \BibitemOpen
  \bibfield  {author} {\bibinfo {author} {\bibfnamefont {P.}~\bibnamefont
  {Agrawal}}\ and\ \bibinfo {author} {\bibfnamefont {K.}~\bibnamefont {Howe}},\
  }\href {\doibase 10.1007/JHEP12(2018)029} {\bibfield  {journal} {\bibinfo
  {journal} {JHEP}\ }\textbf {\bibinfo {volume} {12}},\ \bibinfo {pages} {029}
  (\bibinfo {year} {2018})},\ \Eprint {http://arxiv.org/abs/1710.04213}
  {arXiv:1710.04213 [hep-ph]} \BibitemShut {NoStop}%
\bibitem [{\citenamefont {Alves}\ and\ \citenamefont
  {Weiner}(2018)}]{Alves:2017avw}%
  \BibitemOpen
  \bibfield  {author} {\bibinfo {author} {\bibfnamefont {D.~S.~M.}\
  \bibnamefont {Alves}}\ and\ \bibinfo {author} {\bibfnamefont
  {N.}~\bibnamefont {Weiner}},\ }\href {\doibase 10.1007/JHEP07(2018)092}
  {\bibfield  {journal} {\bibinfo  {journal} {JHEP}\ }\textbf {\bibinfo
  {volume} {07}},\ \bibinfo {pages} {092} (\bibinfo {year} {2018})},\ \Eprint
  {http://arxiv.org/abs/1710.03764} {arXiv:1710.03764 [hep-ph]} \BibitemShut
  {NoStop}%
\bibitem [{\citenamefont {Krauss}\ and\ \citenamefont
  {Wise}(1986)}]{Krauss:1986bq}%
  \BibitemOpen
  \bibfield  {author} {\bibinfo {author} {\bibfnamefont {L.~M.}\ \bibnamefont
  {Krauss}}\ and\ \bibinfo {author} {\bibfnamefont {M.~B.}\ \bibnamefont
  {Wise}},\ }\href {\doibase 10.1016/0370-2693(86)90201-7} {\bibfield
  {journal} {\bibinfo  {journal} {Phys. Lett.}\ }\textbf {\bibinfo {volume}
  {B176}},\ \bibinfo {pages} {483} (\bibinfo {year} {1986})}\BibitemShut
  {NoStop}%
\bibitem [{\citenamefont {Bardeen}\ \emph {et~al.}(1987)\citenamefont
  {Bardeen}, \citenamefont {Peccei},\ and\ \citenamefont
  {Yanagida}}]{BARDEEN1987401}%
  \BibitemOpen
  \bibfield  {author} {\bibinfo {author} {\bibfnamefont {W.~A.}\ \bibnamefont
  {Bardeen}}, \bibinfo {author} {\bibfnamefont {R.}~\bibnamefont {Peccei}}, \
  and\ \bibinfo {author} {\bibfnamefont {T.}~\bibnamefont {Yanagida}},\ }\href
  {\doibase https://doi.org/10.1016/0550-3213(87)90003-4} {\bibfield  {journal}
  {\bibinfo  {journal} {Nuclear Physics B}\ }\textbf {\bibinfo {volume}
  {279}},\ \bibinfo {pages} {401 } (\bibinfo {year} {1987})}\BibitemShut
  {NoStop}%
\bibitem [{\citenamefont {Aoki}\ \emph {et~al.}(2011)\citenamefont {Aoki} \emph
  {et~al.}}]{PIENU:2011aa}%
  \BibitemOpen
  \bibfield  {author} {\bibinfo {author} {\bibfnamefont {M.}~\bibnamefont
  {Aoki}} \emph {et~al.} (\bibinfo {collaboration} {PIENU}),\ }\href {\doibase
  10.1103/PhysRevD.84.052002} {\bibfield  {journal} {\bibinfo  {journal} {Phys.
  Rev.}\ }\textbf {\bibinfo {volume} {D84}},\ \bibinfo {pages} {052002}
  (\bibinfo {year} {2011})},\ \Eprint {http://arxiv.org/abs/1106.4055}
  {arXiv:1106.4055 [hep-ex]} \BibitemShut {NoStop}%
\bibitem [{\citenamefont {Aguilar-Arevalo}\ \emph {et~al.}(2015)\citenamefont
  {Aguilar-Arevalo} \emph {et~al.}}]{Aguilar-Arevalo:2015cdf}%
  \BibitemOpen
  \bibfield  {author} {\bibinfo {author} {\bibfnamefont {A.}~\bibnamefont
  {Aguilar-Arevalo}} \emph {et~al.} (\bibinfo {collaboration} {PiENu}),\ }\href
  {\doibase 10.1103/PhysRevLett.115.071801} {\bibfield  {journal} {\bibinfo
  {journal} {Phys. Rev. Lett.}\ }\textbf {\bibinfo {volume} {115}},\ \bibinfo
  {pages} {071801} (\bibinfo {year} {2015})},\ \Eprint
  {http://arxiv.org/abs/1506.05845} {arXiv:1506.05845 [hep-ex]} \BibitemShut
  {NoStop}%
\bibitem [{\citenamefont {Tanabashi}\ \emph {et~al.}(2018)\citenamefont
  {Tanabashi} \emph {et~al.}}]{PhysRevD.98.030001}%
  \BibitemOpen
  \bibfield  {author} {\bibinfo {author} {\bibfnamefont {M.}~\bibnamefont
  {Tanabashi}} \emph {et~al.} (\bibinfo {collaboration} {Particle Data
  Group}),\ }\href {\doibase 10.1103/PhysRevD.98.030001} {\bibfield  {journal}
  {\bibinfo  {journal} {Phys. Rev. D}\ }\textbf {\bibinfo {volume} {98}},\
  \bibinfo {pages} {030001} (\bibinfo {year} {2018})}\BibitemShut {NoStop}%
\bibitem [{\citenamefont {Aguilar-Arevalo}\ \emph {et~al.}(2018)\citenamefont
  {Aguilar-Arevalo} \emph {et~al.}}]{Aguilar-Arevalo:2017vlf}%
  \BibitemOpen
  \bibfield  {author} {\bibinfo {author} {\bibfnamefont {A.}~\bibnamefont
  {Aguilar-Arevalo}} \emph {et~al.} (\bibinfo {collaboration} {PIENU}),\ }\href
  {\doibase 10.1103/PhysRevD.97.072012} {\bibfield  {journal} {\bibinfo
  {journal} {Phys. Rev.}\ }\textbf {\bibinfo {volume} {D97}},\ \bibinfo {pages}
  {072012} (\bibinfo {year} {2018})},\ \Eprint
  {http://arxiv.org/abs/1712.03275} {arXiv:1712.03275 [hep-ex]} \BibitemShut
  {NoStop}%
\bibitem [{\citenamefont {Aguilar-Arevalo}\ \emph {et~al.}(2019)\citenamefont
  {Aguilar-Arevalo} \emph {et~al.}}]{Aguilar-Arevalo:2019owf}%
  \BibitemOpen
  \bibfield  {author} {\bibinfo {author} {\bibfnamefont {A.}~\bibnamefont
  {Aguilar-Arevalo}} \emph {et~al.} (\bibinfo {collaboration} {PIENU}),\
  }\href@noop {} {\  (\bibinfo {year} {2019})},\ \Eprint
  {http://arxiv.org/abs/1904.03269} {arXiv:1904.03269 [hep-ex]} \BibitemShut
  {NoStop}%
\bibitem [{\citenamefont {Picciotto}\ \emph {et~al.}(1988)\citenamefont
  {Picciotto}, \citenamefont {Ahmad}, \citenamefont {Britton}, \citenamefont
  {Bryman}, \citenamefont {Clifford}, \citenamefont {Kitching}, \citenamefont
  {Kuno}, \citenamefont {Macdonald}, \citenamefont {Numao}, \citenamefont
  {Olin}, \citenamefont {Poutissou}, \citenamefont {Summhammer},\ and\
  \citenamefont {Dixit}}]{PhysRevD.37.1131}%
  \BibitemOpen
  \bibfield  {author} {\bibinfo {author} {\bibfnamefont {C.~E.}\ \bibnamefont
  {Picciotto}}, \bibinfo {author} {\bibfnamefont {S.}~\bibnamefont {Ahmad}},
  \bibinfo {author} {\bibfnamefont {D.~I.}\ \bibnamefont {Britton}}, \bibinfo
  {author} {\bibfnamefont {D.~A.}\ \bibnamefont {Bryman}}, \bibinfo {author}
  {\bibfnamefont {E.~T.~H.}\ \bibnamefont {Clifford}}, \bibinfo {author}
  {\bibfnamefont {P.}~\bibnamefont {Kitching}}, \bibinfo {author}
  {\bibfnamefont {Y.}~\bibnamefont {Kuno}}, \bibinfo {author} {\bibfnamefont
  {J.~A.}\ \bibnamefont {Macdonald}}, \bibinfo {author} {\bibfnamefont
  {T.}~\bibnamefont {Numao}}, \bibinfo {author} {\bibfnamefont
  {A.}~\bibnamefont {Olin}}, \bibinfo {author} {\bibfnamefont {J.-M.}\
  \bibnamefont {Poutissou}}, \bibinfo {author} {\bibfnamefont {J.}~\bibnamefont
  {Summhammer}}, \ and\ \bibinfo {author} {\bibfnamefont {M.~S.}\ \bibnamefont
  {Dixit}},\ }\href {\doibase 10.1103/PhysRevD.37.1131} {\bibfield  {journal}
  {\bibinfo  {journal} {Phys. Rev. D}\ }\textbf {\bibinfo {volume} {37}},\
  \bibinfo {pages} {1131} (\bibinfo {year} {1988})}\BibitemShut {NoStop}%
\bibitem [{\citenamefont {Pocanic}\ \emph {et~al.}(2004)\citenamefont {Pocanic}
  \emph {et~al.}}]{Pocanic:2003pf}%
  \BibitemOpen
  \bibfield  {author} {\bibinfo {author} {\bibfnamefont {D.}~\bibnamefont
  {Pocanic}} \emph {et~al.},\ }\href {\doibase 10.1103/PhysRevLett.93.181803}
  {\bibfield  {journal} {\bibinfo  {journal} {Phys. Rev. Lett.}\ }\textbf
  {\bibinfo {volume} {93}},\ \bibinfo {pages} {181803} (\bibinfo {year}
  {2004})},\ \Eprint {http://arxiv.org/abs/hep-ex/0312030}
  {arXiv:hep-ex/0312030 [hep-ex]} \BibitemShut {NoStop}%
\bibitem [{\citenamefont {Bj{\"o}rkeroth}\ \emph {et~al.}(2018)\citenamefont
  {Bj{\"o}rkeroth}, \citenamefont {Chun},\ and\ \citenamefont
  {King}}]{Bjorkeroth:2018dzu}%
  \BibitemOpen
  \bibfield  {author} {\bibinfo {author} {\bibfnamefont {F.}~\bibnamefont
  {Bj{\"o}rkeroth}}, \bibinfo {author} {\bibfnamefont {E.~J.}\ \bibnamefont
  {Chun}}, \ and\ \bibinfo {author} {\bibfnamefont {S.~F.}\ \bibnamefont
  {King}},\ }\href {\doibase 10.1007/JHEP08(2018)117} {\bibfield  {journal}
  {\bibinfo  {journal} {JHEP}\ }\textbf {\bibinfo {volume} {08}},\ \bibinfo
  {pages} {117} (\bibinfo {year} {2018})},\ \Eprint
  {http://arxiv.org/abs/1806.00660} {arXiv:1806.00660 [hep-ph]} \BibitemShut
  {NoStop}%
\bibitem [{\citenamefont {Wise}(1981)}]{Wise:1980ux}%
  \BibitemOpen
  \bibfield  {author} {\bibinfo {author} {\bibfnamefont {M.~B.}\ \bibnamefont
  {Wise}},\ }\href {\doibase 10.1016/0370-2693(81)90684-5} {\bibfield
  {journal} {\bibinfo  {journal} {Phys. Lett.}\ }\textbf {\bibinfo {volume}
  {103B}},\ \bibinfo {pages} {121} (\bibinfo {year} {1981})}\BibitemShut
  {NoStop}%
\bibitem [{\citenamefont {Frere}\ \emph {et~al.}(1981)\citenamefont {Frere},
  \citenamefont {Vermaseren},\ and\ \citenamefont {Gavela}}]{Frere:1981cc}%
  \BibitemOpen
  \bibfield  {author} {\bibinfo {author} {\bibfnamefont {J.~M.}\ \bibnamefont
  {Frere}}, \bibinfo {author} {\bibfnamefont {J.~A.~M.}\ \bibnamefont
  {Vermaseren}}, \ and\ \bibinfo {author} {\bibfnamefont {M.~B.}\ \bibnamefont
  {Gavela}},\ }\href {\doibase 10.1016/0370-2693(81)90686-9} {\bibfield
  {journal} {\bibinfo  {journal} {Phys. Lett.}\ }\textbf {\bibinfo {volume}
  {103B}},\ \bibinfo {pages} {129} (\bibinfo {year} {1981})}\BibitemShut
  {NoStop}%
\bibitem [{\citenamefont {Hall}\ and\ \citenamefont
  {Wise}(1981)}]{Hall:1981bc}%
  \BibitemOpen
  \bibfield  {author} {\bibinfo {author} {\bibfnamefont {L.~J.}\ \bibnamefont
  {Hall}}\ and\ \bibinfo {author} {\bibfnamefont {M.~B.}\ \bibnamefont
  {Wise}},\ }\href {\doibase 10.1016/0550-3213(81)90469-7} {\bibfield
  {journal} {\bibinfo  {journal} {Nucl. Phys.}\ }\textbf {\bibinfo {volume}
  {B187}},\ \bibinfo {pages} {397} (\bibinfo {year} {1981})}\BibitemShut
  {NoStop}%
\bibitem [{\citenamefont {Bauer}\ \emph {et~al.}(2017)\citenamefont {Bauer},
  \citenamefont {Neubert},\ and\ \citenamefont {Thamm}}]{Bauer:2017ris}%
  \BibitemOpen
  \bibfield  {author} {\bibinfo {author} {\bibfnamefont {M.}~\bibnamefont
  {Bauer}}, \bibinfo {author} {\bibfnamefont {M.}~\bibnamefont {Neubert}}, \
  and\ \bibinfo {author} {\bibfnamefont {A.}~\bibnamefont {Thamm}},\ }\href
  {\doibase 10.1007/JHEP12(2017)044} {\bibfield  {journal} {\bibinfo  {journal}
  {JHEP}\ }\textbf {\bibinfo {volume} {12}},\ \bibinfo {pages} {044} (\bibinfo
  {year} {2017})},\ \Eprint {http://arxiv.org/abs/1708.00443} {arXiv:1708.00443
  [hep-ph]} \BibitemShut {NoStop}%
\bibitem [{\citenamefont {Sirlin}(1978)}]{RevModPhys.50.573}%
  \BibitemOpen
  \bibfield  {author} {\bibinfo {author} {\bibfnamefont {A.}~\bibnamefont
  {Sirlin}},\ }\href {\doibase 10.1103/RevModPhys.50.573} {\bibfield  {journal}
  {\bibinfo  {journal} {Rev. Mod. Phys.}\ }\textbf {\bibinfo {volume} {50}},\
  \bibinfo {pages} {573} (\bibinfo {year} {1978})}\BibitemShut {NoStop}%
\bibitem [{\citenamefont {Cowan}\ \emph {et~al.}(1986)\citenamefont {Cowan},
  \citenamefont {Backe}, \citenamefont {Bethge}, \citenamefont {Bokemeyer},
  \citenamefont {Folger}, \citenamefont {Greenberg}, \citenamefont {Sakaguchi},
  \citenamefont {Schwalm}, \citenamefont {Schweppe}, \citenamefont {Stiebing},\
  and\ \citenamefont {Vincent}}]{PhysRevLett.56.444}%
  \BibitemOpen
  \bibfield  {author} {\bibinfo {author} {\bibfnamefont {T.}~\bibnamefont
  {Cowan}}, \bibinfo {author} {\bibfnamefont {H.}~\bibnamefont {Backe}},
  \bibinfo {author} {\bibfnamefont {K.}~\bibnamefont {Bethge}}, \bibinfo
  {author} {\bibfnamefont {H.}~\bibnamefont {Bokemeyer}}, \bibinfo {author}
  {\bibfnamefont {H.}~\bibnamefont {Folger}}, \bibinfo {author} {\bibfnamefont
  {J.~S.}\ \bibnamefont {Greenberg}}, \bibinfo {author} {\bibfnamefont
  {K.}~\bibnamefont {Sakaguchi}}, \bibinfo {author} {\bibfnamefont
  {D.}~\bibnamefont {Schwalm}}, \bibinfo {author} {\bibfnamefont
  {J.}~\bibnamefont {Schweppe}}, \bibinfo {author} {\bibfnamefont {K.~E.}\
  \bibnamefont {Stiebing}}, \ and\ \bibinfo {author} {\bibfnamefont
  {P.}~\bibnamefont {Vincent}},\ }\href {\doibase 10.1103/PhysRevLett.56.444}
  {\bibfield  {journal} {\bibinfo  {journal} {Phys. Rev. Lett.}\ }\textbf
  {\bibinfo {volume} {56}},\ \bibinfo {pages} {444} (\bibinfo {year}
  {1986})}\BibitemShut {NoStop}%
\bibitem [{\citenamefont {Behrends}\ and\ \citenamefont
  {Sirlin}(1960)}]{PhysRevLett.4.186}%
  \BibitemOpen
  \bibfield  {author} {\bibinfo {author} {\bibfnamefont {R.~E.}\ \bibnamefont
  {Behrends}}\ and\ \bibinfo {author} {\bibfnamefont {A.}~\bibnamefont
  {Sirlin}},\ }\href {\doibase 10.1103/PhysRevLett.4.186} {\bibfield  {journal}
  {\bibinfo  {journal} {Phys. Rev. Lett.}\ }\textbf {\bibinfo {volume} {4}},\
  \bibinfo {pages} {186} (\bibinfo {year} {1960})}\BibitemShut {NoStop}%
\bibitem [{\citenamefont {Ademollo}\ and\ \citenamefont
  {Gatto}(1964)}]{PhysRevLett.13.264}%
  \BibitemOpen
  \bibfield  {author} {\bibinfo {author} {\bibfnamefont {M.}~\bibnamefont
  {Ademollo}}\ and\ \bibinfo {author} {\bibfnamefont {R.}~\bibnamefont
  {Gatto}},\ }\href {\doibase 10.1103/PhysRevLett.13.264} {\bibfield  {journal}
  {\bibinfo  {journal} {Phys. Rev. Lett.}\ }\textbf {\bibinfo {volume} {13}},\
  \bibinfo {pages} {264} (\bibinfo {year} {1964})}\BibitemShut {NoStop}%
\bibitem [{\citenamefont {Leutwyler}\ and\ \citenamefont
  {Roos}(1984)}]{Leutwyler1984}%
  \BibitemOpen
  \bibfield  {author} {\bibinfo {author} {\bibfnamefont {H.}~\bibnamefont
  {Leutwyler}}\ and\ \bibinfo {author} {\bibfnamefont {M.}~\bibnamefont
  {Roos}},\ }\href {\doibase 10.1007/BF01571961} {\bibfield  {journal}
  {\bibinfo  {journal} {Zeitschrift f{\"u}r Physik C Particles and Fields}\
  }\textbf {\bibinfo {volume} {25}},\ \bibinfo {pages} {91} (\bibinfo {year}
  {1984})}\BibitemShut {NoStop}%
\bibitem [{\citenamefont {Bourrely}\ \emph {et~al.}(1981)\citenamefont
  {Bourrely}, \citenamefont {Machet},\ and\ \citenamefont
  {de~Rafael}}]{BOURRELY1981157}%
  \BibitemOpen
  \bibfield  {author} {\bibinfo {author} {\bibfnamefont {C.}~\bibnamefont
  {Bourrely}}, \bibinfo {author} {\bibfnamefont {B.}~\bibnamefont {Machet}}, \
  and\ \bibinfo {author} {\bibfnamefont {E.}~\bibnamefont {de~Rafael}},\ }\href
  {\doibase https://doi.org/10.1016/0550-3213(81)90086-9} {\bibfield  {journal}
  {\bibinfo  {journal} {Nuclear Physics B}\ }\textbf {\bibinfo {volume}
  {189}},\ \bibinfo {pages} {157 } (\bibinfo {year} {1981})}\BibitemShut
  {NoStop}%
\bibitem [{\citenamefont {Berlin}\ \emph {et~al.}(2018)\citenamefont {Berlin},
  \citenamefont {Gori}, \citenamefont {Schuster},\ and\ \citenamefont
  {Toro}}]{Berlin:2018pwi}%
  \BibitemOpen
  \bibfield  {author} {\bibinfo {author} {\bibfnamefont {A.}~\bibnamefont
  {Berlin}}, \bibinfo {author} {\bibfnamefont {S.}~\bibnamefont {Gori}},
  \bibinfo {author} {\bibfnamefont {P.}~\bibnamefont {Schuster}}, \ and\
  \bibinfo {author} {\bibfnamefont {N.}~\bibnamefont {Toro}},\ }\href {\doibase
  10.1103/PhysRevD.98.035011} {\bibfield  {journal} {\bibinfo  {journal} {Phys.
  Rev.}\ }\textbf {\bibinfo {volume} {D98}},\ \bibinfo {pages} {035011}
  (\bibinfo {year} {2018})},\ \Eprint {http://arxiv.org/abs/1804.00661}
  {arXiv:1804.00661 [hep-ph]} \BibitemShut {NoStop}%
\bibitem [{\citenamefont {Dolan}\ \emph {et~al.}(2017)\citenamefont {Dolan},
  \citenamefont {Ferber}, \citenamefont {Hearty}, \citenamefont {Kahlhoefer},\
  and\ \citenamefont {Schmidt-Hoberg}}]{Dolan:2017osp}%
  \BibitemOpen
  \bibfield  {author} {\bibinfo {author} {\bibfnamefont {M.~J.}\ \bibnamefont
  {Dolan}}, \bibinfo {author} {\bibfnamefont {T.}~\bibnamefont {Ferber}},
  \bibinfo {author} {\bibfnamefont {C.}~\bibnamefont {Hearty}}, \bibinfo
  {author} {\bibfnamefont {F.}~\bibnamefont {Kahlhoefer}}, \ and\ \bibinfo
  {author} {\bibfnamefont {K.}~\bibnamefont {Schmidt-Hoberg}},\ }\href
  {\doibase 10.1007/JHEP12(2017)094} {\bibfield  {journal} {\bibinfo  {journal}
  {JHEP}\ }\textbf {\bibinfo {volume} {12}},\ \bibinfo {pages} {094} (\bibinfo
  {year} {2017})},\ \Eprint {http://arxiv.org/abs/1709.00009} {arXiv:1709.00009
  [hep-ph]} \BibitemShut {NoStop}%
\bibitem [{\citenamefont {Aloni}\ \emph {et~al.}(2019)\citenamefont {Aloni},
  \citenamefont {Fanelli}, \citenamefont {Soreq},\ and\ \citenamefont
  {Williams}}]{Aloni:2019ruo}%
  \BibitemOpen
  \bibfield  {author} {\bibinfo {author} {\bibfnamefont {D.}~\bibnamefont
  {Aloni}}, \bibinfo {author} {\bibfnamefont {C.}~\bibnamefont {Fanelli}},
  \bibinfo {author} {\bibfnamefont {Y.}~\bibnamefont {Soreq}}, \ and\ \bibinfo
  {author} {\bibfnamefont {M.}~\bibnamefont {Williams}},\ }\href@noop {} {\
  (\bibinfo {year} {2019})},\ \Eprint {http://arxiv.org/abs/1903.03586}
  {arXiv:1903.03586 [hep-ph]} \BibitemShut {NoStop}%
\bibitem [{Mim(2015)}]{Mimasu:2014nea}%
  \BibitemOpen
  \href {\doibase 10.1007/JHEP06(2015)173} {\bibfield  {journal} {\bibinfo
  {journal} {JHEP}\ }\textbf {\bibinfo {volume} {06}},\ \bibinfo {pages} {173}
  (\bibinfo {year} {2015})},\ \Eprint {http://arxiv.org/abs/1409.4792}
  {arXiv:1409.4792 [hep-ph]} \BibitemShut {NoStop}%
\bibitem [{\citenamefont {Jaeckel}\ and\ \citenamefont
  {Spannowsky}(2016)}]{Jaeckel:2015jla}%
  \BibitemOpen
  \bibfield  {author} {\bibinfo {author} {\bibfnamefont {J.}~\bibnamefont
  {Jaeckel}}\ and\ \bibinfo {author} {\bibfnamefont {M.}~\bibnamefont
  {Spannowsky}},\ }\href {\doibase 10.1016/j.physletb.2015.12.037} {\bibfield
  {journal} {\bibinfo  {journal} {Phys. Lett.}\ }\textbf {\bibinfo {volume}
  {B753}},\ \bibinfo {pages} {482} (\bibinfo {year} {2016})},\ \Eprint
  {http://arxiv.org/abs/1509.00476} {arXiv:1509.00476 [hep-ph]} \BibitemShut
  {NoStop}%
\bibitem [{\citenamefont {Banerjee}\ \emph {et~al.}(2018)\citenamefont
  {Banerjee} \emph {et~al.}}]{Banerjee:2017hhz}%
  \BibitemOpen
  \bibfield  {author} {\bibinfo {author} {\bibfnamefont {D.}~\bibnamefont
  {Banerjee}} \emph {et~al.} (\bibinfo {collaboration} {NA64}),\ }\href
  {\doibase 10.1103/PhysRevD.97.072002} {\bibfield  {journal} {\bibinfo
  {journal} {Phys. Rev.}\ }\textbf {\bibinfo {volume} {D97}},\ \bibinfo {pages}
  {072002} (\bibinfo {year} {2018})},\ \Eprint
  {http://arxiv.org/abs/1710.00971} {arXiv:1710.00971 [hep-ex]} \BibitemShut
  {NoStop}%
\bibitem [{Ake(2018)}]{Akesson:2018vlm}%
  \BibitemOpen
  \href@noop {} {\  (\bibinfo {year} {2018})},\ \Eprint
  {http://arxiv.org/abs/1808.05219} {arXiv:1808.05219 [hep-ex]} \BibitemShut
  {NoStop}%
\bibitem [{\citenamefont {del Amo~Sanchez}\ \emph {et~al.}(2011)\citenamefont
  {del Amo~Sanchez} \emph {et~al.}}]{delAmoSanchez:2010ac}%
  \BibitemOpen
  \bibfield  {author} {\bibinfo {author} {\bibfnamefont {P.}~\bibnamefont {del
  Amo~Sanchez}} \emph {et~al.} (\bibinfo {collaboration} {BaBar}),\ }\href
  {\doibase 10.1103/PhysRevLett.107.021804} {\bibfield  {journal} {\bibinfo
  {journal} {Phys. Rev. Lett.}\ }\textbf {\bibinfo {volume} {107}},\ \bibinfo
  {pages} {021804} (\bibinfo {year} {2011})},\ \Eprint
  {http://arxiv.org/abs/1007.4646} {arXiv:1007.4646 [hep-ex]} \BibitemShut
  {NoStop}%
\bibitem [{\citenamefont {Bergsma}\ \emph {et~al.}(1985)\citenamefont {Bergsma}
  \emph {et~al.}}]{Bergsma:1985qz}%
  \BibitemOpen
  \bibfield  {author} {\bibinfo {author} {\bibfnamefont {F.}~\bibnamefont
  {Bergsma}} \emph {et~al.} (\bibinfo {collaboration} {CHARM}),\ }\href
  {\doibase 10.1016/0370-2693(85)90400-9} {\bibfield  {journal} {\bibinfo
  {journal} {Phys. Lett.}\ }\textbf {\bibinfo {volume} {157B}},\ \bibinfo
  {pages} {458} (\bibinfo {year} {1985})}\BibitemShut {NoStop}%
\bibitem [{\citenamefont {Blumlein}\ \emph {et~al.}(1991)\citenamefont
  {Blumlein} \emph {et~al.}}]{Blumlein:1990ay}%
  \BibitemOpen
  \bibfield  {author} {\bibinfo {author} {\bibfnamefont {J.}~\bibnamefont
  {Blumlein}} \emph {et~al.},\ }\href {\doibase 10.1007/BF01548556} {\bibfield
  {journal} {\bibinfo  {journal} {Z. Phys.}\ }\textbf {\bibinfo {volume}
  {C51}},\ \bibinfo {pages} {341} (\bibinfo {year} {1991})}\BibitemShut
  {NoStop}%
\bibitem [{\citenamefont {Blumlein}\ \emph {et~al.}(1992)\citenamefont
  {Blumlein} \emph {et~al.}}]{Blumlein:1991xh}%
  \BibitemOpen
  \bibfield  {author} {\bibinfo {author} {\bibfnamefont {J.}~\bibnamefont
  {Blumlein}} \emph {et~al.},\ }\href {\doibase 10.1142/S0217751X9200171X}
  {\bibfield  {journal} {\bibinfo  {journal} {Int. J. Mod. Phys.}\ }\textbf
  {\bibinfo {volume} {A7}},\ \bibinfo {pages} {3835} (\bibinfo {year}
  {1992})}\BibitemShut {NoStop}%
\bibitem [{Dob(2016)}]{Dobrich:2015jyk}%
  \BibitemOpen
  \href {\doibase 10.1007/JHEP02(2016)018} {\bibfield  {journal} {\bibinfo
  {journal} {JHEP}\ }\textbf {\bibinfo {volume} {02}},\ \bibinfo {pages} {018}
  (\bibinfo {year} {2016})},\ \bibinfo {note} {[JHEP02,018(2016)]},\ \Eprint
  {http://arxiv.org/abs/1512.03069} {arXiv:1512.03069 [hep-ph]} \BibitemShut
  {NoStop}%
\bibitem [{\citenamefont {Bjorken}\ \emph {et~al.}(1988)\citenamefont
  {Bjorken}, \citenamefont {Ecklund}, \citenamefont {Nelson}, \citenamefont
  {Abashian}, \citenamefont {Church}, \citenamefont {Lu}, \citenamefont {Mo},
  \citenamefont {Nunamaker},\ and\ \citenamefont {Rassmann}}]{Bjorken:1988as}%
  \BibitemOpen
  \bibfield  {author} {\bibinfo {author} {\bibfnamefont {J.~D.}\ \bibnamefont
  {Bjorken}}, \bibinfo {author} {\bibfnamefont {S.}~\bibnamefont {Ecklund}},
  \bibinfo {author} {\bibfnamefont {W.~R.}\ \bibnamefont {Nelson}}, \bibinfo
  {author} {\bibfnamefont {A.}~\bibnamefont {Abashian}}, \bibinfo {author}
  {\bibfnamefont {C.}~\bibnamefont {Church}}, \bibinfo {author} {\bibfnamefont
  {B.}~\bibnamefont {Lu}}, \bibinfo {author} {\bibfnamefont {L.~W.}\
  \bibnamefont {Mo}}, \bibinfo {author} {\bibfnamefont {T.~A.}\ \bibnamefont
  {Nunamaker}}, \ and\ \bibinfo {author} {\bibfnamefont {P.}~\bibnamefont
  {Rassmann}},\ }\href {\doibase 10.1103/PhysRevD.38.3375} {\bibfield
  {journal} {\bibinfo  {journal} {Phys. Rev.}\ }\textbf {\bibinfo {volume}
  {D38}},\ \bibinfo {pages} {3375} (\bibinfo {year} {1988})}\BibitemShut
  {NoStop}%
\bibitem [{Dob(2018)}]{Dobrich:2017gcm}%
  \BibitemOpen
  \bibfield  {booktitle} {\emph {\bibinfo {booktitle} {{Proceedings of the
  PHOTON-2017 Conference: CERN, Geneva, Switzerland, May 22-26, 2017}}},\
  }\href {\doibase 10.23727/CERN-Proceedings-2018-001.253} {\bibfield
  {journal} {\bibinfo  {journal} {CERN Proc.}\ }\textbf {\bibinfo {volume}
  {1}},\ \bibinfo {pages} {253} (\bibinfo {year} {2018})},\ \Eprint
  {http://arxiv.org/abs/1708.05776} {arXiv:1708.05776 [hep-ph]} \BibitemShut
  {NoStop}%
\bibitem [{\citenamefont {Beacham}\ \emph {et~al.}(2019)\citenamefont {Beacham}
  \emph {et~al.}}]{Beacham:2019nyx}%
  \BibitemOpen
  \bibfield  {author} {\bibinfo {author} {\bibfnamefont {J.}~\bibnamefont
  {Beacham}} \emph {et~al.},\ }\href@noop {} {\  (\bibinfo {year} {2019})},\
  \Eprint {http://arxiv.org/abs/1901.09966} {arXiv:1901.09966 [hep-ex]}
  \BibitemShut {NoStop}%
\bibitem [{\citenamefont {Feng}\ \emph {et~al.}(2018)\citenamefont {Feng},
  \citenamefont {Galon}, \citenamefont {Kling},\ and\ \citenamefont
  {Trojanowski}}]{Feng:2018noy}%
  \BibitemOpen
  \bibfield  {author} {\bibinfo {author} {\bibfnamefont {J.~L.}\ \bibnamefont
  {Feng}}, \bibinfo {author} {\bibfnamefont {I.}~\bibnamefont {Galon}},
  \bibinfo {author} {\bibfnamefont {F.}~\bibnamefont {Kling}}, \ and\ \bibinfo
  {author} {\bibfnamefont {S.}~\bibnamefont {Trojanowski}},\ }\href {\doibase
  10.1103/PhysRevD.98.055021} {\bibfield  {journal} {\bibinfo  {journal} {Phys.
  Rev.}\ }\textbf {\bibinfo {volume} {D98}},\ \bibinfo {pages} {055021}
  (\bibinfo {year} {2018})},\ \Eprint {http://arxiv.org/abs/1806.02348}
  {arXiv:1806.02348 [hep-ph]} \BibitemShut {NoStop}%
\bibitem [{\citenamefont {Alekhin}\ \emph {et~al.}(2016)\citenamefont {Alekhin}
  \emph {et~al.}}]{Alekhin:2015byh}%
  \BibitemOpen
  \bibfield  {author} {\bibinfo {author} {\bibfnamefont {S.}~\bibnamefont
  {Alekhin}} \emph {et~al.},\ }\href {\doibase 10.1088/0034-4885/79/12/124201}
  {\bibfield  {journal} {\bibinfo  {journal} {Rept. Prog. Phys.}\ }\textbf
  {\bibinfo {volume} {79}},\ \bibinfo {pages} {124201} (\bibinfo {year}
  {2016})},\ \Eprint {http://arxiv.org/abs/1504.04855} {arXiv:1504.04855
  [hep-ph]} \BibitemShut {NoStop}%
\bibitem [{\citenamefont {Frlez}\ \emph {et~al.}(2004)\citenamefont {Frlez}
  \emph {et~al.}}]{Frlez:2003vg}%
  \BibitemOpen
  \bibfield  {author} {\bibinfo {author} {\bibfnamefont {E.}~\bibnamefont
  {Frlez}} \emph {et~al.},\ }\href {\doibase 10.1016/j.nima.2004.03.137}
  {\bibfield  {journal} {\bibinfo  {journal} {Nucl. Instrum. Meth.}\ }\textbf
  {\bibinfo {volume} {A526}},\ \bibinfo {pages} {300} (\bibinfo {year}
  {2004})},\ \Eprint {http://arxiv.org/abs/hep-ex/0312017}
  {arXiv:hep-ex/0312017 [hep-ex]} \BibitemShut {NoStop}%
\end{thebibliography}
